 \documentclass{aa}
 \usepackage{graphicx}
 \usepackage{times}
 \usepackage{color}
\begin{document}
 
\def\Msun{M_{\sun}}
\def\rsun{R_{\sun}}
\def\Mp{M_{\rm p}}
\def\Rp{R_{\rm P}}
\def\rj{R_{\rm out}}
\def\Mjup{M_{\rm J}}
\def\rjup{R_{\rm J}}
\def\ri{r_{\rm in}}
\def\cp{circum-planetary }
\def\Cp{Circum-planetary }
\def\cs{circum-stellar }
\def\Cs{Circum-stellar }
\newbox\grsign \setbox\grsign=\hbox{$>$}
\newdimen\grdimen \grdimen=\ht\grsign
\newbox\laxbox \newbox\gaxbox
\setbox\gaxbox=\hbox{\raise.5ex\hbox{$>$}\llap
     {\lower.5ex\hbox{$\sim$}}}\ht1=\grdimen\dp1=0pt
\setbox\laxbox=\hbox{\raise.5ex\hbox{$<$}\llap
     {\lower.5ex\hbox{$\sim$}}}\ht2=\grdimen\dp2=0pt
\def\gax{$\mathrel{\copy\gaxbox}$}
\def\lax{$\mathrel{\copy\laxbox}$}

\title{Magnetically driven outflows from Jovian circum-planetary 
       accretion disks}
\author{Christian Fendt}
\institute{
        Institut f\"ur Physik, Universit\"at Potsdam, 
        Am Neuen Palais 10,
        D-14469 Potsdam, Germany; \email{cfendt@aip.de} 
          }
\date{Received <date>; accepted <date> }
\authorrunning{Ch.~Fendt}
\titlerunning{Circum-planetary disk outflows} 
\abstract{
We discuss the possibility to launch an outflow from the close vicinity
of a protoplanetary core considering a model scenario where the protoplanet
surrounded by a circum-{\em planetary} accretion disk is located in a
circum-{\em stellar} disk.
For the circum-planetary disk accretion rate we assume
$ \dot{M}_{\rm cp} \simeq 6\times10^{-5} \Mjup{\rm yr}^{-1}$
implying peak disk temperatures of about 2000\,K.
The estimated disk ionization degree and Reynolds number allow for a
sufficient coupling between the disk matter and the magnetic field.
We find that the surface magnetic field strength of the protoplanet is
probably not more than 10\,G,
indicating that the {\em global} planetary magnetosphere
is dominated by the circum-planetary disk magnetic field of \lax 50\,G.
The existence of a gap between circum-planetary disk and planet 
seems to be unlikely.
The estimated field strength and mass flow rates allow for asymptotic
outflow velocities of \gax $60\,{\rm km\,s}^{-1}$.
The overall outflow geometry will be governed by the orbital radius,
resembling a hollow tube or cone perpendicular the disk.
The length of the outflow built up during one orbital period is
about 100\,AU, depending on the outflow velocity.
Outflows from circum-planetary disks may be visible in shock excited 
emission lines along a tube of diameter of the orbital radius and 
thickness of about 100 protoplanetary radii. 
We derive particle densities of 3000\,${\rm cm}^{-3}$ in this 
layer. 
Energetically, protoplanetary outflows cannot survive the interaction
with a {\em protostellar} outflow.
Due to the efficient angular momentum removal by the outflow,
we expect the protoplanetary outflow to influence the early planet
angular momentum evolution.
If this is true, 
planets which have produced an outflow in earlier times will rotate 
slower at later times.
The mass evolution of the planet is, however, 
hardly affected as the outflow mass loss rate will be small compared 
to the mass accumulated by the protoplanetary core.
}
\maketitle

\section{Introduction}
With the discovery of {\em extrasolar planets} during the last decade
the scientific interest in planet formation has been increased 
substantially.
In particular, with the help of the computational power existing today,
it has become possible to perform numerical simulations of a circumstellar
accretion disk containing and building up an orbiting protoplanetary core
(e.g. Kley \cite{kley99};
Bryden et al.~\cite{bryd99};
Lubow et al.~\cite{lubo99};
Kley et al.~\cite{kley01};
D'Angelo et al.~\cite{dang02};
Tanigawa \& Watanabe \cite{tani02};
D'Angelo et al.~\cite{dang03a}, \cite{dang03b};
Bate et al.~\cite{bate03}).
Although differing in certain aspects
(spatial resolution of the region close to the protoplanet,
number of dimensions treated)
these simulations have provided us with the same general results.
In particular, 
the simulations show how tidal interaction between the protoplanet and the
disk material opens up a {\em gap} in the \cs accretion disk along the
orbit of the planetary core. 
Mass accretion in the \cs disk, however, continues across the gap.
The \cs disk material entering the Roche lobe of the protoplanet becomes
captured and finally accreted by the protoplanetary core.
The simulations also demonstrate how the accretion stream initially
affected by strong shocks waves eventually builds up a 
{\em \cp} sub-disk in almost Keplerian rotation close to the planet.


On the other hand, in the context of {\em astrophysical jets} it is well
established that outflow formation is causally connected to the presence
of an {\em accretion disk} and strong {\em magnetic fields}
(see 
Blandford \& Payne \cite{blan82};
Pudritz \& Norman \cite{pudr86};
Camenzind \cite{came90};
Shu et al. \cite{shu94}).
This statement holds for a wide range in outflow energy and spatial
scale -- for young stellar objects, microquasars and active galactic nuclei.
Observations as well as theoretical models strongly suggest that
astrophysical sources of outflows are (highly) magnetized.
The magnetic field is responsible for acceleration and collimation
and for lifting the matter from the disk into the outflow.

In the scenario of planet formation within a {\cs accretion disk,} 
the presence of a magnetic field can be expected as well.
A central protostellar dipolar magnetic field of 1000\,G surface field
strength may provide only about 20$\mu$G at 5\,AU distance.
This (weak) field, however, may act as a seed field for dynamo action
taking place in the \cs accretion disk, in the protoplanetary core,
or in the \cp disk.
Most of the present-day Solar system planets carry a substantial
magnetic field.
We expect these fields to be present also during the phase of planet 
formation.

Considering 
(i) the numerically established existence of accretion disks around
protoplanetary cores as a natural feature during the planet formation 
phase 
and (ii) the feasibility of a large-scale magnetic field in the 
protoplanetary environment,
the question arises whether the combination of these two factors
may lead to the launch of an outflow 
similar to the phenomenon observed on much larger spatial and
energetic scales.
Up to date, the literature on this topic is rather rare.
To our knowledge, 
the possible outflow activity from Jupiter-sized protoplanets
has been investigated only by a single paper
(Quillen \& Trilling \cite{quil98}).
Applying stationary models of magnetohydrodynamic outflow formation
developed for {\em protostars}, 
these authors came to the conclusion that protoplanetary outflows might
indeed exist with velocities $\simeq 20\,{\rm km\,s}^{-1}$ and 
mass loss rates $\simeq 10^{-8}$ Jupiter masses per year.

In the present paper we further investigate the feasibility of
launching outflows from accretion disks around protoplanets.
The paper is organized as follows. 
Section 2 summarizes the basic ideas of magnetohydrodynamically
driven outflows and their possible application to protoplanets.
We then discuss the circum-planetary accretion disk. Section 3.1
is devoted to numerical simulations of circum-planetary disk in
the literature. 
We estimate the temperature (Sect.~3.2) and
ionization state (Sect.~3.3) of such disks.
The next section considers the magnetic field of protoplanet (Sect.~4.1),
and circum-planetary disk (Sect.~4.2), and the expected MHD properties
of the outflow (Sect.~4.4, 4.5).
Based on the estimated properties of the system we discuss
the parameter space for MHD driven outflows from circum-planetary
disks and their observational appearance (Sect.~5).
We conclude the paper with our summary (Sect.~6).

We will consider a complex model scenario consisting of a number of 
constituents.
In order to label their parameters, we use the following notation.
Parameters of the planetary core have the subscript ``p'', 
those of the outflow the subscript ``out'', 
those of the \cp disk the subscript ``cp'', 
those of the central star the subscript ``$\star$''. 
Parameters of the present-day Jupiter are denoted by the 
subscript ``J''. 
Further subscripts will be defined in the text.
Variables without subscript denote general quantities.

\section{Magnetohydrodynamic (MHD) formation of outflows}
\subsection{The MHD model of astrophysical outflow formation}
We first outline briefly the essential features of MHD formation
of outflows.
The generally accepted ``standard'' model understands the outflow 
as {\em magnetically} driven 
(Blandford \& Payne \cite{blan82};
Pudritz \& Norman \cite{pudr86}; 
Camenzind \cite{came90}; 
Shu et al. \cite{shu94}).
Following current theoretical models the outflow originates in the
innermost part of a magnetized ``star''-disk system 
(the ``star'' can be a young stellar object or a collapsed object).
Whether the outflow magnetic field is primarily anchored in the accretion
disk
or in the surface of the central object, is not yet clear.
The fact that outflows are observed also from sources containing
a central black hole implies that the key magnetic field component 
is indeed provided by the surrounding accretion disk.
The existence of non-relativistic outflows implies that the relativistic
character of the source of an outflow cannot be a major constraint for its
launching.

We summarize the MHD formation of outflows process as follows.
\begin{itemize}
\item
The {\em magnetic field} necessary for launching can be generated by 
dynamo action in the ``star''-disk system or advected from the interstellar
medium.
\item
A fraction of the accreting plasma is lifted up to higher altitudes by magnetic 
forces.
The matter couples to the large-scale magnetic field, launching a slow wind.
\item
The matter is then accelerated, first by {\em magneto-centrifugal}
forces, then, further out, by {\em Lorentz forces}.
Poynting flux is converted to kinetic energy.
\item
Inertial forces lead to a ``bending'' of the poloidal field lines
(i.e. the induction of a {\em toroidal magnetic field}).
The toroidal field tension {\em collimates} the outflow into a 
narrow jet.
\item
The plasma velocity subsequently exceeds the speed of the {\em MHD waves}.
The asymptotic, super-magnetosonic flow is ballistic and 
{\em causally decoupled} from outer boundary conditions.
\item
At the point where the outflow front meets the interstellar medium, 
a {\em bow shock} develops. 
\end{itemize}
It has been discussed in the literature that outflows can only be formed
in a geometrical configuration underlying a certain
{\em degree of axisymmetry}
(Fendt \& Zinnecker \cite{fend98}).
In the case of a circumplanetary disk-outflow-system, an axisymmetric
alignment is indicated as the strong differential rotation of the
\cp disk would control the dynamo mechanism to provide magnetic fields
(in the disk or in the planet) aligned with the rotational axis.

Recent numerical simulations confirm this scenario of outflow formation
which has been developed over many years mainly by stationary MHD 
models.
Axisymmetric MHD simulations have shown the self-collimating property
of MHD flows from {\em rotating disks} (Ouyed \& Pudritz \cite{ouye97}) also 
under the de-collimating effect of a turbulent magnetic diffusivity
(Fendt \& Cemeljic \cite{fend02}).
A {\em stellar wind} dominated disk-outflow does not collimate on the 
spatial scales considered in the simulations 
(Fendt \& Elstner \cite{fend00}).
Three-dimensional simulations of collimating disk winds prove that 
strong non-axisymmetric perturbations at the base of the outflow indeed 
lead to growing MHD instabilities eventually disrupting the outflow flow
(Ouyed et al.~\cite{ouye03}).

\subsection{The question of protoplanetary outflows}
In this section we identify the critical points for
MHD outflow formation in the context of protoplanetary outflows.
First, one has to discuss the existence of a \cp accretion disk 
and their parameters 
(accretion rate, temperature and degree of ionization).
\Cp accretion disks have not yet been investigated in detail 
theoretically or numerically.
Therefore, in our estimates we will apply the following approach.
We expect the outflow to be launched in the inner part of the 
{\em \cp} disk.
For this inner sub-disk, 
we assume a mass flow rate as determined by 
the numerical simulations of the {\em \cs} accretion disk,
and treat the disk structure assuming a standard, thin 
Shakura-Sunyaev disk. 
%

In the second step we will derived some estimates on the magnetic field
structure close to the protoplanetary core.
Two scenarios are plausible -- an accretion disk equipartition
magnetic field or the field of the protoplanet itself.

In general, the numerical simulations of the accretion stream
around the protoplanet do {\em not} show any 
indication for outflow activity around the protoplanet.
However, we know that astrophysical outflow are launched magnetically.
Therefore, we cannot expect at all to detect such a process in a purely
hydrodynamic treatment, which, in addition, does not resolve the region
close to the planet.
Radiation driven or gas pressure winds can be launched from a very hot
accretion disk, however, such a scenario is unlikely for protoplanets 
where relatively cool disk temperatures are required to allow for solid
dust grain condensation (Bryden et al.~\cite{bryd99}).

\section{The \cp accretion disk}
\subsection{Numerical evidence for \cp accretion disks}
During the last years, a number of papers have been published considering
time-dependent hydrodynamic simulations of a \cs accretion disk which 
is enclosing an orbiting protoplanetary core
(Bryden et al.~\cite{bryd99};
Lubow et al.~\cite{lubo99};
Nelson et al.~\cite{nels00}; 
Kley et al.~\cite{kley01}; 
D'Angelo et al.~\cite{dang02};
Bate et al.~\cite{bate03}).
In summary,
all these simulations have demonstrated how the tidal interaction between 
the protoplanet and the viscous \cs accretion disk opens a gap along 
the orbit.
The feasibility of such a scenario has been predicted some decades ago
(Lin \& Papaloizou \cite{lin86,lin93}; 
Artymowicz \& Lubow \cite{arty96}).
A similar dynamical evolution is known from simulations of circum-binary 
accretion disks 
(Artymowicz \& Lubow \cite{arty94})

Essentially, accretion of matter within the \cs disk continues across 
the gap.
That part of the material which spirals around the Roche lobe of the
planetary core will then enter the Roche lobe and finally be captured by 
the planetary gravitational potential.
The spiraling matter generates a density wave propagating into the
\cs disk.
The path of the material approaching the Roche lobe follows more and more a
\cp orbital motion representing the feature of a {\em \cp accretion disk}.
Depending on the mass of the planetary core the gas orbiting within 
the {\em \cp disk} reaches different rotational regimes.
The simulations show that the orbital velocity profile can deviate from a
Keplerian rotation by up to 10\% primarily due to the smoothing of the
gravitational potential
(D'Angelo et al.\cite{dang02}).
An almost circular orbital symmetry is reached close to the central planet.
On the other hand, at smaller radii the deviation from the Keplerian velocity
is larger.

The governing parameter for our studies is the {\em accretion rate} of
the \cp disk.
In general, we find a good agreement in the above mentioned papers 
concerning this value. 
Kley et al.~(\cite{kley01}) give an accretion rate towards the planetary
core of 
\begin{equation}
\label{eq_m_dot}
\dot{M}_{\rm cp} \simeq 6\times10^{-5} \Mjup{\rm yr}^{-1}, 
\end{equation}
a value which is confirmed also by three-dimensional simulations
(D'Angelo et al.~\cite{dang03a}).
The exact value depends on certain model assumptions for the \cs disk
as the disk mass distributed along the orbit 
($3.5\times10^{-3}\Msun$ between 2 and 13\,AU), 
the disk height ($h(r)=0.05\,r$) 
and the viscosity parameter ($\alpha \simeq 4\times 10^{-3}$).
Lubow et al.~(\cite{lubo99}) give accretion rates typical for their
simulations of 
$\dot{M}_{\rm cp} \simeq 4.5\times 10^{-5}\Mjup{\rm yr}^{-1}$ for similar
disk parameters as Kley et al.~(\cite{kley01}).
Bryden et al.~(\cite{bryd99}) obtain accretion rates between 
$\dot{M}_{\rm cp} \simeq 10^{-5}\Mjup{\rm yr}^{-1}$ and  
$\dot{M}_{\rm cp} \simeq 3\times10^{-4}\Mjup{\rm yr}^{-1}$ 
which are, as expected, larger in the case of a higher disk viscosity.
In the present paper, we will in general refer to the model parameters of 
Kley et al.~(\cite{kley01}) and D'Angelo et al. (\cite{dang02}) and
apply their numerically derived accretion rate of 
$6\!\times\!10^{-5} \Mjup{\rm yr}^{-1}$.
This value is substantially higher than the value estimated by 
Quillen \& Trilling (\cite{quil98}).

A difficulty with applying these numerical simulations to the 
\cp accretion flow is their lack of sufficient numerical resolution in
this region.
In particular, this holds for the innermost part of the \cp disk where we 
expect the outflow to be launched.
The interesting length scales here are the planetary gravitational radius 
$r_{\rm H} = D_{\rm p} (\Mp / 3 M_{\star})^{1/3}$ (Hill radius),
%
\begin{equation}
\label{eq_r_hill}
r_{\rm H} = 508\rjup \left(\frac{D_{\rm p}}{5.2{\rm AU}}\right)
             \left(\frac{M_p}{M_J}\right)^{1/3}
             \left(\frac{M_{\star}}{\Msun}\right)^{-1/3},
\end{equation}
the radius of the planetary core of presumably several $\rjup$,
and the orbital radius $D_{\rm p}$.

Recent simulations by Tanigawa \& Watanabe (\cite{tani02}), 
particularly considering the gas accretion flow close to the protoplanet,
apply a numerical grid extending from $0.025\,r_{\rm H}$ to 
$12\,r_{\rm H}$ resolving
$0.005\,r_{\rm H}$ close to the planet.
For parameters similar to Eq.~\ref{eq_r_hill} this corresponds to about
$\simeq 20\rjup$ 
and covers reasonably well the region where we expect outflow formation to
happen.
Due to a different setup -- no gap in the \cs disk along the orbit --
the accretion rate towards the planetary core is higher by factor of
hundred compared to the literature 
(Lubow et al.~\cite{lubo99}, D'Angelo et al.~\cite{dang02}).
Compensating for this effect the authors confirm values for the
re-scaled surface densities and accretion rates similar to 
Eq.~\ref{eq_m_dot}. 
%
Further, Tanigawa \& Watanabe deal with an isothermal disk, and, 
thus, cannot deliver the temperature profile in the disk. 
The authors claim that the main features of the flow dynamics are 
roughly the same for adiabatic and isothermal simulations
(with a factor 3 difference in the accretion rate).

Due to lack of numerical resolution close to the planet,
the simulations published so far give little information about the
{\em structure} of the \cp disk (e.g. Kley \cite{kley01}).
Three-dimensional simulations are essential to determine the 
scale height $(h/r)$ of the sub-disk.
Further, the disk structure also depends on the number value for the 
viscosity parameter 
which is generally taken constant throughout the whole computational
domain 
(in particular, it is assumed to be the same for \cs and \cp disk).
So far, 
the simulations have shown that the scale height of the sub-disk
is
{\em much smaller} compared to the main disk due to the gravitational
potential of the planet (Bate et al.~\cite{bate03}).

Preliminary results of very high resolution numerical simulations 
(Ciecielag et al.~\cite{ciec00}) suggest that the spiral density wave
launched in the outer regions of the \cp disk may propagate into the inner
disk regions.
However, simulations by Bate et al.~(\cite{bate03}) indicate that the 
strong spiral shocks observed in 2D simulations are greatly diminished
in 3D simulations.
In recent simulations by D'Angelo (\cite{dang03}) 
the \cp disk becomes almost axisymmetric in the region within $20\,\rjup$
from the planet.
In the end, it is up to future high-resolution \cp disk simulations to
show whether the simple assumption of an axisymmetric, thin Keplerian disk must
be revised.

In summary,
numerical hydrodynamic simulations of {\em \cs} accretion disks give clear
evidence for the existence of \cp disks in almost Keplerian rotation.
%
In order to understand further the accretion process towards the
protoplanetary core,
the numerical simulation of the {\em \cp} accretion disk itself is
essential.
No such work has yet been presented in the literature.
Therefore, we have to rely on 
estimates obtained by scaling the standard accretion disk parameters
to protoplanetary parameters taking into account the results of the above 
mentioned \cs disk simulations as model constraints.
In the following we will consider a {\em thin disk} with accretion rate as
given in Eq.~\ref{eq_m_dot}.

\subsection{\Cp disk temperature and density}
\label{sec_t_disk}
Only matter of a sufficiently hot and, thus, sufficiently ionized 
accretion disk can couple to and interact with a magnetic field,
eventually leading to the launch of winds or outflows.
Therefore, we first need an estimate for the disk temperature profile.
As discussed above, in order to estimate the temperature profile of
the \cp accretion disk, 
we will consider the limit of a standard (thin, Shakura-Sunyaev) disk
(see also Nelson \& Benz \cite{nels03})
with an accretion rate Eq.~\ref{eq_m_dot} provided by the numerical 
{\em \cs} disk simulations.

A first estimate of the \cp disk temperature can be derived from
the well known equation for the effective surface temperature of a 
geometrically thin disk 
surrounding a central body of mass $M$ and an accretion rate $\dot{M}$ 
as a function of radius $r$ is
$$
T_{\rm s}(r) = \left(\frac{3}{8\pi\sigma}\frac{G M \dot{M}}{r^3}\right)^{1/4}
$$
with the Stefan-Boltzmann constant $\sigma$
(see e.g. Pringle \cite{prin81}).
For Jupiter-like parameters and a \cp disk accretion rate as in 
Eq.~\ref{eq_m_dot} we obtain
\begin{equation}
\label{eq_t_disk}
T_{\rm s}(r) = 1280\,{\rm K}
\left(\frac{M}{\Mjup}\!\right)^{\!\frac{1}{4}}
\!\left(\!\frac{\dot{M}_{\rm cp}}{6\!\times\!10^{-5} \Mjup\,{\rm yr}^{-1}} 
\!\right)^{\!\!\frac{1}{4}}
\!\left(\!\frac{r}{10\,\rjup}\!\right)^{\!-\frac{3}{4}}
\end{equation}
Interestingly, we see that the ``power'' of the disk $\sim M\dot{M}$ normalized
to the mass of the central object is similar for the \cp and the \cs case.
However, the \cp disk extends a factor $(\rsun/\rjup)$ closer to the origin of
the gravitational potential.
Therefore, a high temperature and luminosity can be expected in the inner disk.

The disk temperature as estimated in Eq.~\ref{eq_t_disk} is substantially
larger than that of Quillen \& Trilling (\cite{quil98}).
This is due to the much higher accretion rate we consider as motivated 
by the numerical simulations.
The temperature derived in Eq.~\ref{eq_t_disk} is a measure for the disk 
{\em surface} temperature.

Viscous heating of the disk will lead to even higher temperatures.
For comparison, we show the mid-plane temperature distribution in the
case of the $\alpha$-parametrization for the viscosity 
(Shakura \& Sunyaev \cite{shak73}) in the appendix.
As a result, we find a remarkable similarity between the \cp and the
\cs accretion disks.

A straightforward estimate of the mid-plane disk temperature can be made 
by assuming radiative diffusion in the disk.
%
Then the effective black body surface temperature is related to the central
mid-plane temperature by 
$
T_{\rm c}^4 = (3/8)\tau T^4_{\rm eff}
$
(Hubeny \cite{hube90}).
This gives mid-plane temperatures $\simeq 2-4 $ higher than the
surface temperature and rather insensitive to the vertical optical depth 
$\tau$.
%
%
Our simple estimates yielding a rather high \cp accretion disk temperature
is in agreement also with disk models discussing the formation of the
Galilean satellites (Canup \& Ward \cite{canu02}).

In order to derive a realistic disk internal temperature profile, 
more sophisticated theoretical model calculations are needed 
taking into account radiation transfer and a realistic description
of the opacity.
Only few such models are available for protostellar accretion disks
(e.g. D'Alessio et al.~\cite{dale98}, Malbet et al.~\cite{malb01},
Dullemond et al.~\cite{dull02}),
none for \cp disks.
For the example of T\,Tauri star accretion disks, it has been found
(Malbet et al.~\cite{malb01})
that the central disk temperature is a factor ten higher than the 
surface temperature within a disk radius of about 10\,AU.
Malbet et al. have pointed out that the other important heating source 
for the disk besides viscosity, namely the radiation of the central 
4000\,K star has not been taking into account in their model.
D'Alessio et al.~(\cite{dale98}) consider stellar irradiation
as the main heating source of the disk upper layers. 
Similarly, Tanigawa \& Watanabe (\cite{tani02}) have argued that the
temperature profile of a \cp sub-disk within the Hill radius is
substantially affected by heating of the protoplanetary core, in particular
in the late stages of planetary growth.
For our estimates, 
we do not consider the detailed vertical structure of the disk and the
effect of irradiation.
Our simple temperature estimate must be therefore considered as a lower 
limit.

Recent 2D thermo-hydrodynamic simulations (D'Angelo et al.~\cite{dang03b})
were able to resolve the temperature profile close to the planet with
reasonable resolution.
These simulations take into account radiative heating and cooling as 
well as a sophisticated opacity model.
The authors find a peak mid-plane temperature of 1200 - 1600\,K 
at radii of $r=5$ - $10\,\rjup$ from the protoplanet (D'Angelo 2003).
Essentially, these temperatures are somewhat below the values given
by our simple ansatz of a standard disk.
This agrees with models of \cs disks taking into account radiative
transfer and which give in general a lower mid-plane temperatures compared
to the treatment without radiative transfer (Dullemond et al.~\cite{dull02}).

A very high disk temperature might be problematic as potentially leading to
a disk evaporation as discussed for disks around collapsed objects 
(e.g. Meyer \& Meyer-Hofmeister \cite{meye94}).
Disk evaporation would affect just that (innermost) part of the
\cp disk where we expect the origin of the outflow.
However, 
due to the fact that the \cp disk reaches deeper into the
gravitational potential of the central body,
its $\alpha$-disk scale height is less 
compared to the case of a \cs disk (see eq.~\ref{eq_hr}).
This, and also the general similarity to the \cs disk, 
indicates that the \cp disk does not blow up and remains a thin disk
close to the planet.

\subsection{The \cp disk ionization}
\label{sec_i_disk}
A minimum degree of ionization is necessary to couple the matter to
the magnetic field sufficiently. 
In the case of protostellar accretion disks several sources of 
disk ionization have been identified,
(i) 
thermal ionization,
(ii) 
ionization by cosmic ray particles,
and (iii)
ionization by non-thermal X-rays.
As discussed above, the temperature and density profiles are 
comparable in \cs and \cp accretion disks.
Since the \cp disk is located within the \cs disk, it is straight forward
to  assume that also the ionization sources are the same,
indicating that also the the degree of disk ionization will be similar in
both cases.

To determine the ionization degree of a protostellar accretion disk is
a complex enterprise and depends on a number of unknown quantities
(see e.g. Gammie \cite{gamm96}, D'Alessio et al.~\cite{dale98}, 
Fromang et al.~\cite{from02}).
In our case, we are interested in the innermost part of the 
{\em \cp} accretion disk, 
where the launch of an outflow may take place.
In this region the disk is hot and dominated by thermal ionization.
The degree of ionization is then
\begin{eqnarray}
\label{eq_x_ion}
\xi & = & 4.2\times 10^{-13} \left(\frac{a}{10^{-7}}\right)^{1/2}
             \left(\frac{T_{\rm c}}{10^3{\rm K}}\right)^{3/4} \nonumber \\
  & \, & \quad\quad\quad\quad
 \cdot \left(\frac{n_{\rm n}}{10^{15}{\rm cm}^{-3}}\right)^{-1/2}
 \frac{\exp(-25188\,T^{-1})}{1.15 \times 10^{-11}}
\end{eqnarray}
with the neutral particle density $n_{\rm n}$ and $a$ the
abundance of K atoms relative to Hydrogen
(see Fromang et al.~\cite{from02} and references therein).
Thermal ionization is dominant for mid-plane temperatures \gax$10^3$K. 
Applying our previous estimates for the temperature profile in \cp disks,
such temperatures can be expected for radii
$r$\,\lax\,$65\,\rjup$.
Considering density values as given in Eq.~(A.2) we find an ionization
degree above $10^{-11}$ for $r$\,\lax\,$65\,\rjup$.

Note that the degree of thermal ionization of the {\em \cs disk} at 
the orbital radius of $5.2\,$AU is much lower (Eq.~\ref{eq_x_ion})
and negligible.
However, cosmic ray ionization may still provide ionization levels up
to $10^{-11}$ at these radii
(Gammie\footnote{There are still large uncertainties in the
ionization rate along the disk surface, mostly connected to
the possible impact of the stellar wind} 
\cite{gamm96}, see Eq.~\ref{eq_re_m})
and may therefore also contribute to the \cp disk ionization.
The vertical profile of the ionization level can only be calculated
numerically. 
Model simulations of \cs disks indicate that lower disk layers may 
actually be decoupled from the magnetic field 
(Fromang et al.~\cite{from02}).
This has already been discussed by Gammie (\cite{gamm96}) who
pointed out sufficient coupling between matter and field is essential
in order to excite the disk {\em turbulence} via the magneto-rotational
instability (Balbus \& Hawley \cite{balb91}).
The results by Gammie further indicate a {\em two-layered accretion} 
structure in protostellar accretion disks for radii $> 0.1\,$AU.
Regions in which the magnetic Reynolds number is $Re_{\rm M} < 1$ are 
magnetically ``inactive''.

It is interesting to work out these numbers for a {\em \cp} disk.
The magnetic Reynolds number is related to a certain degree of
ionization. For the case of a \cp disk, we obtain
\begin{equation}
Re_{\rm M} = 7.6\times 10^{15} \xi\,\alpha^{1/2} 
\left(\frac{r}{\rjup}\right)^{3/2}\!\!
\left(\frac{T}{500\,{\rm K}}\right)\!
\left(\frac{M}{\Mjup}\right)^{-1/2}
\label{eq_re_m}
\end{equation}
Using the number values derived above, we see that in the inner part of
the disk a sufficient ionization level $\xi$ \gax $10^{-15}$ is always 
guaranteed providing a good coupling of matter and magnetic field
as an essential condition in order to launch an outflow.

\section{The magnetic environment of the protoplanet}
We know from other astrophysical sources that outflow formation
is basically a {\em magnetic} phenomenon -- outflows are
launched, accelerated and collimated by magnetic forces.
Similar to the case of protostars and protostellar outflows,
two model scenarios for the magnetic field structure around the 
protoplanetary core seem to be feasible.
One is that of a magnetic field distribution dominated by the
{\em protoplanetary magnetosphere}.
In the other, the {\em accretion disk} magnetic field is the main
driver for the outflow.
In the following we will elaborate in greater detail the scenario 
of a magnetohydrodynamic interrelation between the protoplanetary core, 
surrounding accretion disk and a possible outflow.

\subsection{The protoplanetary magnetosphere}
Many of the solar system planets carry a substantial magnetic field, 
in particular the outer gas giants.
Jupiter has a well ordered, strong dipolar magnetic field of about
10\,G aligned with the rotational axis.
It must be mentioned, however, that the detailed structure and orientation 
of the solar system planets differs substantially and the theoretical 
understanding of the underlying (necessarily diverse) dynamo processes
is not yet understood 
(e.g. Connerney \cite{conn93}, Moss \& Brandenburg \cite{moss95}).

So far, we are not aware of any theoretical treatment about the
magnetic field evolution in protoplanets.
Therefore, our considerations of the structure and field strength
of the protoplanetary magnetosphere have to rely on simplifying
estimates. 

Nevertheless, we know from the comparison of stars and protostars
that the large scale magnetic field strength observed in young stars is 
generally higher than the field strength found in main sequence stars.
The main reason is that the 
strength of stellar dynamo action depends on convection (turbulence) 
and rotation 
(see Bouvier \cite{bouv90}, Ghosh \cite{ghos95} and references therein).
These parameters are clearly more pronounced in the early phase of
star formation.
It might be straight forward to assume a similar extrapolation
for the early formation phase of a planet.
This would imply that the large-scale protoplanetary magnetic field has 
been larger than that of a present-day planet.
The seed magnetic field for the protoplanetary dynamo can be provided by the
magnetosphere of the central star and/or the \cs accretion disk where the
planet is embedded in.
What kind of dynamo ($\alpha^2 $ or $\alpha\Omega$) is operating in a
protoplanet is not known.
Similar to protostars a dynamo driven by turbulent convection ($\alpha$) 
is therefore feasible.
Whether, in addition, differential rotation ($\Omega$) plays a role remains
unclear.
It has been shown that fully convective protostars do not rotate
differentially (see below).
There is theoretical indication from the planet formation 
that a giant gaseous protoplanet reaches 
an almost fully convective state within some hundred years 
(Wuchterl et al. \cite{wuch00}).
However, as for protostars, the presence of an accretion disk and the
inferred magnetic coupling between disk and planet will supply a substantial
amount of differential rotation.
Thus, both kinds of dynamo action can be expected,
an $\alpha\Omega$-dynamo producing an aligned axisymmetric dipole,
or
an $\alpha^2$-dynamo with a non-axisymmetric field structure.
Also a quadrupolar field structure are feasible, in particular for high
Taylor numbers (Grote et al.~\cite{grot99}).

The maximum field strength which could be generated by a dynamo can be 
estimated considering the fact that convection must be super-Alfv\'enic 
(Camenzind \cite{came97};
see also K\"uker \& R\"udiger \cite{kuek99}).
The Alfv\'en speed for this critical (mean) field strength is 
$v_{\rm A} \simeq B_{\rm cr} / \sqrt{4\pi \overline{\rho}}$
and has to be compared to the convective velocity defined by the convective 
turnover time scale $\tau_{\rm con}$.
The convective velocity $v_{\rm c}$ can be derived from the energy transport
in the protoplanet. 
Considering the convective radiation flux in the limit of mixing length
theory, the luminosity of the protoplanet is 
$L_{\rm p} = \epsilon_{\rm c} \Mp v_{\rm c}^3 / \Rp $.

Theoretical models of the planetary evolution have shown that the luminosity 
of a young planet can be very high, 
$L_{\rm p} \simeq 10^{-4} L_{\sun}$ (Burrows et al.~\cite{burr01}).
Therefore, we estimate the mean convective velocity in a protoplanet as
\begin{equation}
v_{\rm c} = 14\,{\rm m\,s}^{-1}
\left(\frac{\epsilon_{\rm c}}{30}\right)^{\!-\frac{1}{3}}
\!\!\left(\frac{L_{\rm p}}{10^{-4} L_{\sun}}\right)^{\!\!\frac{1}{3}} 
\!\!\left(\frac{\Mp}{\Mjup}\right)^{\!\!-\frac{1}{3}}
\!\!\left(\frac{\Rp}{3\rjup}\right)^{\!\!\frac{1}{3}}\!\!\!\!.
\end{equation}
Comparison to the mean Alfv\'en speed in the planet provides the critical
magnetic field strength for a hypothetical protoplanetary dynamo,
\begin{equation}
B_{\rm cr} = 10\,{\rm G}
\left(\frac{\epsilon_{\rm c}}{30}\right)^{\!-\frac{1}{3}}
\!\left(\!\frac{M_{\rm T}}{0.2}\!\right)^{\!-1}
\!\!\left(\!\frac{L_{\rm p}}{10^{-4} L_{\sun}}\!\right)^{\!\frac{1}{3}}
\!\!\left(\frac{\Mp}{\Mjup}\right)^{\!\frac{1}{6}}
\!\!\left(\frac{\Rp}{3\rjup}\!\right)^{\!-\frac{7}{6}}\!\!\!\!\!
\end{equation}
with the Alfv\'en Mach number of the convective motion 
$M_{\rm T} \equiv v_{\rm con}/v_{\rm A}$.
A higher field strength would imply a magnetic quenching mechanism 
suppressing the convective motion (and the turbulence) which drives 
the dynamo. 

This estimate of an upper limit for the magnetic field strength in fully
convective protoplanets is interesting in two respects.
Firstly, the derived field strength is not higher than the field strength 
observed in present-day solar system planets. 
If this similarity is true, it indicates a substantial difference between
the magnetic evolution of protostars and protoplanets.
Secondly, this field strength is well below the equipartition field strength
in the \cp accretion disk (see below).
Therefore, we believe that, in difference to the case of protostars, 
the protoplanetary magnetic field cannot dominate the dynamics of the
system even close to the planetary core.

It has been found that a good indicator for the level of magnetic 
activity of a star with rotational period $P$ is the Rossby number 
$Ro = P/\tau_{\rm con} $ (see Pizzolato et al.~\cite{pizz03}).
If we adapt this relation for the case of a protoplanet we find 
\begin{eqnarray}
Ro & = & 2.3\times10^{-3}
\left(\frac{\Omega_p}{\Omega_K(\Rp)}\right)^{-1}
\left(\frac{\epsilon_{\rm con}}{30}\right)^{1/3}
\!\left(\frac{L_{\rm p}}{10^{-4} L_{\sun}}\right)^{1/3} \nonumber \\
& \, & 
\quad\quad\quad\quad\quad\cdot
\!\left(\frac{\Mp}{10^{-3} M_{\sun}}\right)^{-5/6}
\!\left(\frac{\Rp}{3\,\rjup}\right)^{5/6}.
\end{eqnarray}
Compared to protostars (Camenzind \cite{came97}),
this value is somewhat higher (factor 3/2),
indicating that the protoplanet can exhibit a high magnetic
activity as the central protostar.
However, the high level of activity is probably compensated by the weaker
energy output due to the relatively low surface magnetic field strength.

So far we have neglected the effect of dissipative processes for the
dynamo mechanism.
Diffusive and viscous effects will certainly determine the dynamical
evolution of the system. 
%
These effects are widely unstudied, and probably not yet fully understood 
(see e.g. Blackman \cite{blac03}, Vishniac et al.~\cite{vish03}).
Therefore, 
in the following we will discuss some general characteristics of
the protostellar dynamo by estimating dimensionless dynamo numbers
(see Camenzind \cite{came97}, K\"uker \& R\"udiger \cite{kuek99})
in the framework of the mean field dynamo theory.

The first point to note is that the low Rossby number $Ro <<1$ indicates
the protoplanet as a fast rotator.
Thus, the scale of the dynamo $\alpha$-effect, the $\alpha$-dynamo parameter
$\alpha_{\rm D}$, is given by
the convective velocity $\alpha_{\rm D} \simeq v_{\rm c}$.
Another interesting question is whether the $\alpha$ or $\Omega$-effect
plays the major role for the dynamo.
This can be estimated by comparing the magnetic Reynolds numbers
$R_{\alpha} \equiv \alpha_{\rm D} R_{\rm P} / \eta$ and
$R_{\Omega} \equiv \Delta\Omega_{\rm P} R_{\rm P}^2 / \eta$,
where $\eta$ is the turbulent magnetic diffusivity in the
planet and $\Delta\Omega_{\rm P}$ the differential rotation.
As a matter of fact, the latter quantity is quite unknown.
For $\Delta\Omega_{\rm P} \simeq \Omega_{\rm P}$ we find a
maximum ratio
$(R_{\Omega}/R_{\alpha}) \simeq \Omega_{\rm P} R_{\rm P} / v_{\rm c}
\simeq 10^2$.
On the other hand,
it has been shown that fully convective protostars rotate almost 
rigidly (K\"uker \& R\"udiger \cite{kuek97}).
This holds in general for spherical stars without latitudinal gradients
in pressure, density and temperature.
We are attempted to extrapolate these results to fully convective
{\em protoplanetary cores}.
As a consequence, $\Delta\Omega_{\rm P} << \Omega_{\rm P}$ and the
mean field dynamo is dominated by the $alpha$-effect (turbulence).

The $\alpha$-effect becomes more efficient for faster rotators, as  
noted by K\"uker \& R\"udiger (\cite{kuek99}) for protostars.
Also this statement can certainly be generalized from protostars to
protoplanets.
In difference to protostars, the protoplanetary core is colder.
Thus, the magnetic Prandtl number (measuring viscosity in terms of diffusivity)
is generally small. 
For planets we have only $P_{\rm m} = (\nu/\eta) \sim 10^{-5} ... 10^{-2}$ 
(see also Kirk \& Stevenson \cite{kirk87}, Starchencko \& Jones \cite{star02}).
In turn, this implies that the time scale for the turbulent planetary
dynamo is very short.
We expect a rapid increase in field strength during the linear regime up to 
a substantial value,
and later a slow saturation to the maximum field strength 
(non-linear quenching of the dynamo-driving turbulence).

\subsection{Magnetosphere-disk interaction}
We now address the question whether the planetary magnetic field may,
similar to the case of protostars, disrupt the inner part of the disk
forming a gap between the planet and an inner disk radius.
The gas pressure $p$ in a standard disk with an accretion rate $\dot{M}$
and a disk half thickness $h(r)$ around a central body of mass $M$ is 
(neglecting factors of orders of unity)
\begin{equation}
\label{eq_p_disk}
p(r) = 
\frac{1}{\alpha}
\frac{\dot{M}}{4\pi r^2}
\left(\frac{G\,M}{r}\right)^{1/2}
\left(\frac{h}{r}\right)^{-1}
\end{equation}
(Camenzind \cite{came90}; see also R\"udiger et al.~\cite{rued95}).
The quantity $\alpha$ is the usual disk viscosity parameter for
the viscous shear stress $t_{r\phi} = -\alpha\,p$.
For a dipolar magnetic field with $B \sim r^{-3}$, 
pressure equilibrium between the magnetic pressure of the 
protoplanetary magnetosphere and the accretion disk gas pressure
defines the {\em inner disk radius} $\ri$.
In units of the radius of the central object $R$, the inner
disk radius is located at 
\begin{equation}
\label{eq_rin_1}
\frac{\ri}{R} = 
\left(\frac{\alpha}{2}\right)^{2/7}
\left(\frac{G M}{R}\right)^{-1/7}
\left(\frac{h}{r}\right)^{2/7}
\left(B_{\rm s} R \right)^{4/7}
\dot{M}^{-2/7}.
\end{equation}
Here, $B_{\rm s}$ is the magnetic field strength on the surface of
the central body.
If the magnetic pressure is strong enough, it will open up a gap
between the surface of the central body and the disk.
In the case of protostellar magnetospheres, the inner disk radius is
located at about 3 protostellar radii
(Camenzind \cite{came90}, K\"onigl \cite{koen91}).
However, for the case of a \cp accretion disk and a $10\,$G central
dipolar planetary magnetic field the situation would be quite different,
\begin{eqnarray}
\label{eq_rin_2}
\frac{\ri}{\Rp} & = & 0.11 
\alpha^{2/7}
\left(\frac{B_{\rm S}}{10\,{\rm G}}\right)^{4/7}
\left(\frac{\dot{M}_{\rm cp}}{6\times 10^{-5}\,\Mjup\,{\rm yr}^{-1}}
\right)^{-2/7} \\ \nonumber
& \phantom{=} & \quad \quad \cdot
\left(\frac{M}{\Mjup}\right)^{-1/7}
\left(\frac{\Rp}{3\,\rjup}\right)^{5/7}
\left(\frac{h/r}{0.1}\right)^{2/7}.
\end{eqnarray}
Such a low (academic) value for the inner disk radius indicates
that for the accretion rate considered here the dynamical impact of the 
planetary magnetosphere for the disk structure is small.
Only for a $50\,{\rm G}$ surface field strength in combination
with an accretion rate 100 times lower
the inner disk radius moves away from the surface of the planet to
a radius $\ri \simeq 2 \Rp$.
%
%
This estimate is consistent with the results of Quillen \& Trilling 
(\cite{quil98}) who derive $\ri \simeq 1.3 \Rp $.
%
%
Quillen \& Trilling argue that during later evolutionary stages during
the planet formation the accretion rate will decrease, allowing for
a disk disruption. 
However, we like to point out that a lower accretion rate also implies
a lower mass loss rate for the outflow.

The previous estimates demonstrate again how critical the scenario of
protoplanetary outflow formation depends on the \cp accretion rate.
More detailed numerical simulations of the \cp accretion disk evolution
are essential in order to clarify this point.
So far, we have to rely on the simulations presented in the literature
which give a high accretion rate.

Although we have shown that it is unlikely that a protoplanetary magnetic
field can open up a gap in the \cp accretion disk it is interesting to
calculate the {\em co-rotation radius} $r_{\rm co}$ of such a magnetosphere. 
For a rotational period of the protoplanet of $P_{\rm p} = 10^{\rm h}$,
we find
\begin{equation}
r_{\rm co} = 2.25\,\rjup 
\left(\frac{\Mp}{\Mjup}\right)^{1/3}
\left(\frac{P_{\rm p}}{10^{\rm h}}\right)^{2/3},
\end{equation}
which is comparable to its radius.
The ratio $r_{\rm co}/\rjup$ is about a factor 10 larger than for the 
protostellar case.
Note that break-up rotational period for a $\Rp = 3\,\rjup$ ($\Rp = 2\,\rjup$)
protoplanetary core is about $15\,$h ($8\,$h).

As shown above, the \cp disk most probably reaches below the co-rotation
radius, implying an angular velocity at the basis of an outflow 
{\em higher} than the one of the central planet -- in difference
to the case of protostars where the outflow emanates from regions 
outside the co-rotation radius.

It has been claimed in the literature that a stellar magnetic field
sufficiently strong to open up a gap between the star and the disk would
be {\em automatically strong enough}
(e.g. Quillen \& Trilling ) to launch a magnetized wind.
However, one has to accept the fact that there is (yet) no numerical 
proof of such a statement.
Stationary state solutions cannot trace the evolution of an outflow
(Camenzind \cite{came90}; Shu et al.~\cite {shu94}; 
Fendt et al.~\cite{fend95})
and time-dependent simulations of outflow formation either do not consider
a dipolar central field (e.g. Ouyed \& Pudritz \cite{ouye97}) or do
not show evidence for a collimated outflow (Miller \& Stone \cite{mill97};
Goodson et al.~\cite{good97}; Fendt \& Elstner \cite{fend00}).
The fact that strongly magnetized neutron stars and white dwarfs do not
have collimated outflows is another interesting feature in this respect.
On the other hand, outflows are observed in active galactic nuclei emanating
from the accretion disk around a black hole which does not generate 
a magnetic field at all.

\subsection{The \cp disk magnetic field}
Circum-stellar accretion disks are believed to carry a substantial
magnetic field.
Under certain circumstances, this disk field can act as a driver for
the observed outflow activity.
Further, a disk magnetic field of moderate strength plays the essential
role of generating the disk {\em turbulence} by the magneto-rotational
instability (Balbus \& Hawley \cite{balb91}) -- 
which enables the disk angular momentum transport
and eventually allows for accretion of matter at all.
In turn, turbulent motion in a rotating ionized medium may give rise 
to ``mean field'' dynamo action ($\alpha$-effect).

As discussed above (see Sect.~\ref{sec_t_disk} and \ref{sec_i_disk})
there exist strong similarities between \cs and \cp accretion disks
concerning the degree of ionization and the temperature distribution.
In the following we will assume that dynamo action in a \cp disk is 
indeed feasible.
Another possibility to accumulate magnetic flux in the \cp disk is by
advection of ambient field (i.e. the magnetic field induced in the
surrounding \cs disk) within the accretion stream.

From Eq.~\ref{eq_p_disk} we derive a \cp disk equipartition field
strength of
\begin{eqnarray}
\label{eq_beq}
B_{\rm eq}(r) & = & 480\,{\rm G}\,\alpha^{-1/2}
\left(\frac{\dot{M}_{\rm cp}}{6\times 10^{-5}\,\Mjup\,{\rm yr}^{-1}}
\right)^{1/2} 
\!\!\left(\frac{M}{\Mjup}\right)^{1/4}
\\ \nonumber
 \phantom{ B_{\rm eq}} & \, & \quad\quad\quad
\cdot
\left(\frac{h/r}{0.1}\right)^{-1/2}
\left(\frac{r}{10\,\rjup}\right)^{-5/4}.
\end{eqnarray}
This field strength is definitely higher than the surface field strength 
estimated above for a protoplanetary dynamo, 
in particular, since we have to consider a turbulence parameter $\alpha <1$.
We may therefore conclude that a hypothetical protoplanetary outflow
is {\em launched from the disk} and not from the planet itself.
The actual disk magnetic field strength is certainly below the equipartition
field and depends on the detailed properties of the dynamo 
(see e.g. R\"udiger et al.~\cite{rued95}).
In general, the toroidal disk magnetic field strength below the equipartition 
value and the
poloidal field strength one or two orders of magnitude below the toroidal field,
with $B_r \simeq B_z$ depending on the dynamo number.
Adding up, this gives a estimate for the \cp accretion disk
poloidal magnetic field at $r=10\,\rjup$ of about 10 - 50\,G,
if we consider $\alpha \simeq 0.01$.

A disk field strength similar to the equipartition field strength
would lead to disk instabilities.
The disk magnetic field amplified by the magneto-rotational instability
induced turbulence will be saturated by the buoyant escape in vertical
direction leading to the formation of a magnetized disk corona
(Merloni \cite{merl03}). Further, disks with high Poynting flux implying
a relatively strong poloidal magnetic field are favored by  a low
viscosity.
These results obtained for disks around stellar mass black holes may be 
tentatively generalized to other parameters regimes.
Three-dimensional MHD simulations show that a magnetic torus with
initially equipartition field strength will change into a state in which
the global field strength is (1/5) of the equipartition field strength 
(Machida et al.~\cite{mach00}). However, locally, magnetic pressure 
dominated filaments can be found.

If the differential rotation of the disk ($\Omega$-effect) completely dominates 
the magnetic field induction by the dynamo, 
the magnetic disk flux will be negligible,
and the outflow might be launched as a magnetic ``plasma gun''
(see Contopoulos \cite{cont95}).
If the $\alpha$-effect generates a substantial poloidal field component,
we can expect an outflow to be launched similar to the well-known magneto-centrifugal
mechanism (Blandford \& Payne \cite{blan82}).

%

\begin{figure*}
\begin{minipage}{8.5cm}
\includegraphics[width=8cm]{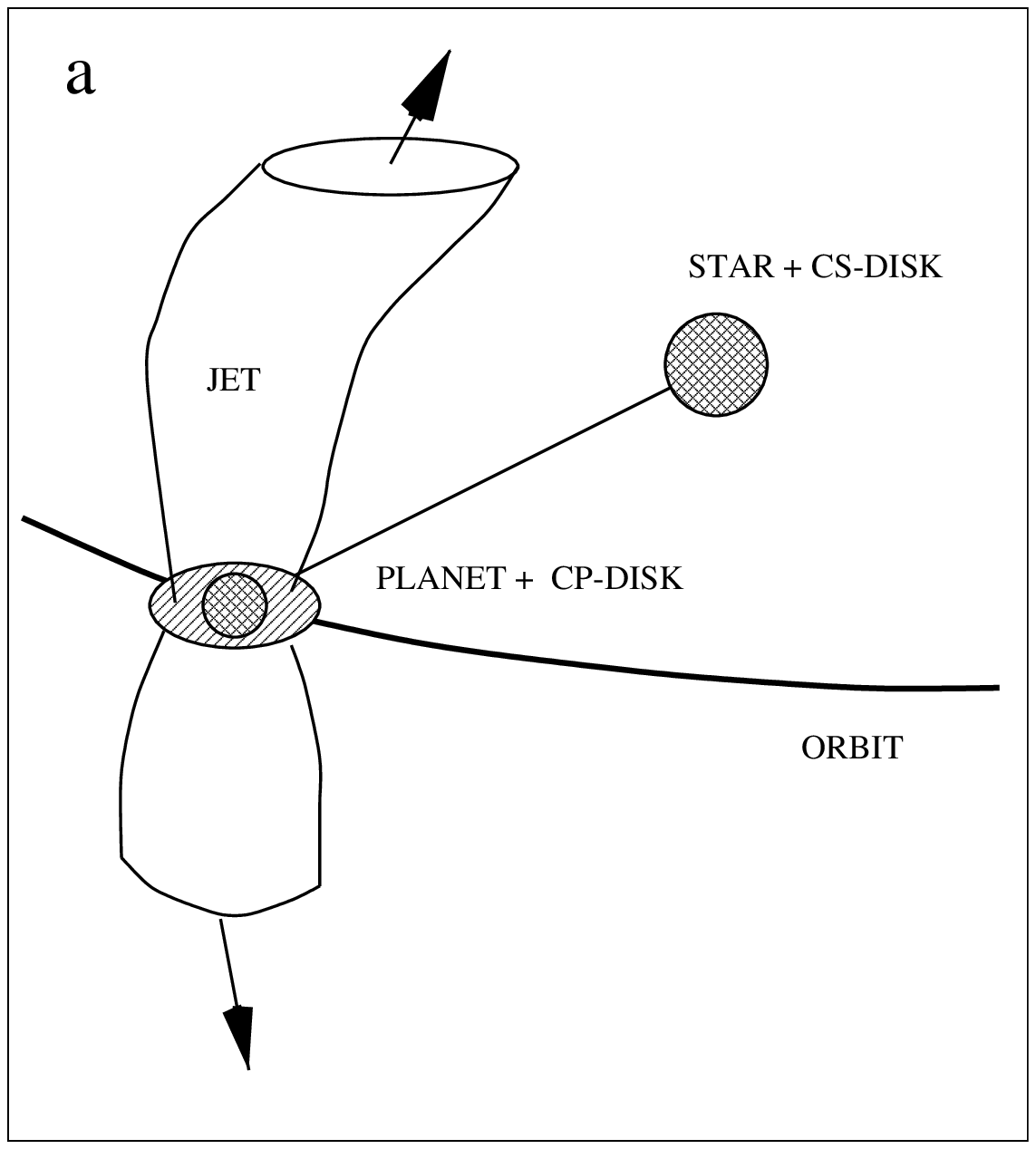}
\end{minipage}
\begin{minipage}{8.5cm}
\includegraphics[width=8cm]{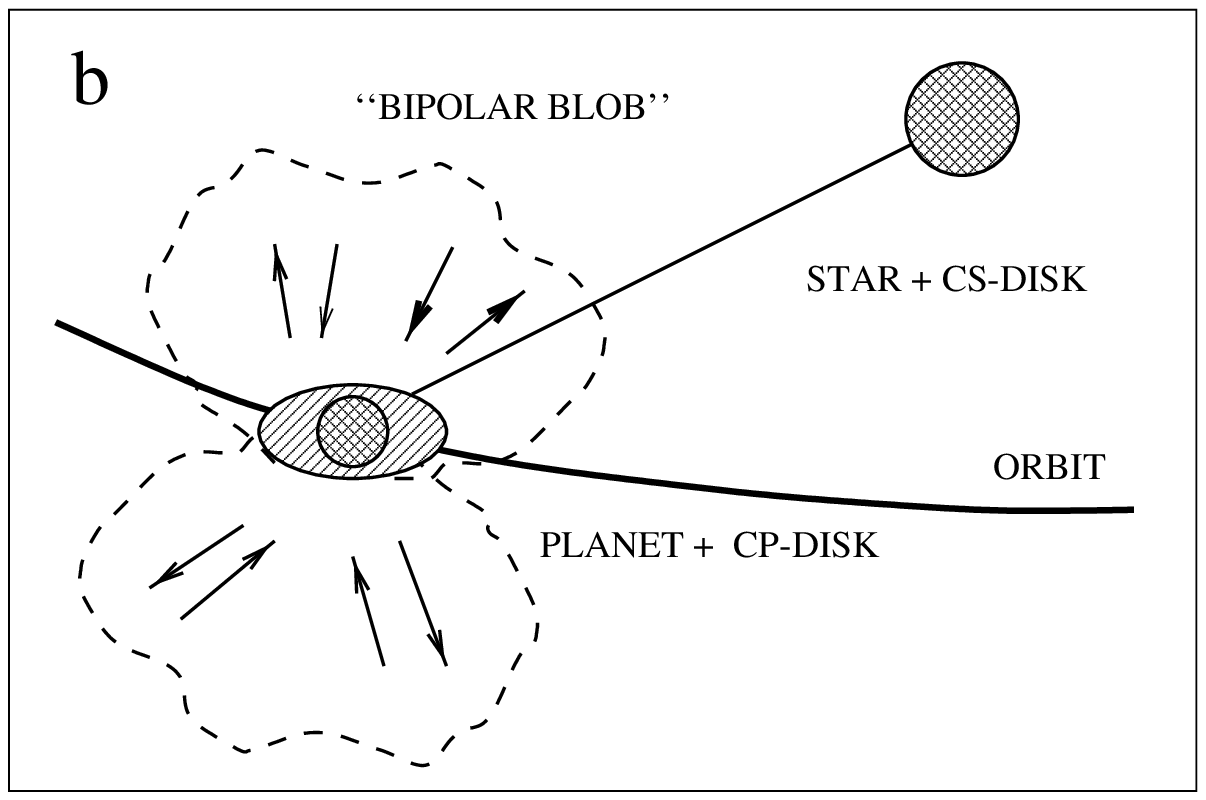}
\end{minipage}
\caption{Model scenario of hypothetical outflows from circum-planetary
accretion disks.
A protoplanet surrounded by a \cp disk (``CP-disk'') is located
within the \cs disk (``CS-disk'').
Depending on the outflow velocity achieved, two different types
of outflow activity is possible -- a large scale outflow, possibly
collimated, or a bipolar blob orbiting with the planet.
}
\end{figure*}

\subsection{Magnetization and asymptotic outflow velocity}
The essential parameter determining the asymptotic speed of the MHD 
outflow, is its magnetization,
\begin{equation}
\sigma = \frac{\Phi^2 \Omega_{\rm F}^2}{c^3 \dot{M}_{\rm out}}
\end{equation}
(Michel \cite{mich69}).
Here, $\Omega_{\rm F}$ is the angular velocity at the foot point
radius $r_F$ of a magnetic field line,
$\Phi\equiv r_F^2 B_F$ refers to the magnetic flux from that area 
and $\dot{M}_{\rm out}$ to the outflow mass loss rate.
For an outflow expanding in spherically {\em radial direction,} Michel 
(\cite{mich69}) derived an analytical relation between the asymptotic flow 
velocity follows and the magnetization,
$v_{\infty} \sim \sigma^{1/3}$ (Michel scaling).
We emphasize the importance of having a rapid rotation ($\Omega_{\rm F}$)
in order to launch a fast outflow.
{\em Faster rotators launch faster outflows}.
It has been shown that the power law interrelation above 
is maintained also for{\em collimating} outflows, 
however with a different power index.
In fact, 
the Michel scaling can be applied for a wide range of astrophysical outflows
from protostellar outflows with
$v_{\rm out} \simeq 500$km/s and $\sigma \simeq 10^{-8}$
to the extragalactic case with
$v_{\rm out} \simeq c$ and $\sigma \simeq 10$
Fendt \& Camenzind (\cite{fend96}).


Applying the Michel scaling for the asymptotic outflow velocity from
\cp accretion disks, we obtain 
\begin{eqnarray}
\label{eq_v_infty}
v_{\infty} & \simeq & 63\,{\rm km\,s}^{-1}
\left(\frac{\Phi_{\rm cp}}{5\times 10^{22}{\rm G\,cm}^2}\right)^{2/3}
\left(\frac{P}{4^{\rm d}}\right)^{-2/3}
 \\ \nonumber
& \, &
\quad\quad\quad
\cdot \left(\frac{\dot{M}_{\rm out}}{10^{-3}\dot{M}_{\rm cp}}\right)^{-1/3}
\left(\frac{\dot{M}_{\rm cp}}{6\times 10^{-5}\,\Mjup\,{\rm yr}}
\right)
^{-1/3}.
\end{eqnarray}
%
Here, we have considered the magnetic flux 
$\Phi_{\rm cp} = 5\times 10^{22}{\rm G\,cm}^2$ corresponding to a poloidal
magnetic field component of $B_{\rm p} \simeq 10\,$G and a Keplerian
period at $r = 10\,\rjup$.
For the outflow mass flow rate we assume 0.1\% of the disk accretion rate, 
a typical value derived in MHD outflow formation models.
For comparison, the asymptotic speed of an outflow launched from
the planetary surface would be $v_{\infty} \simeq 13\,{\rm km\,s}^{-1}$
only, with the magnetic flux $\Phi_{\rm P} = 5\times 10^{20}{\rm G\,cm}^2$
and a rotational period of $10^{\rm h}$ (present-day Jupiter).

The asymptotic outflow velocity as derived in Eq.~\ref{eq_v_infty} 
actually represents a lower limit. 
Taking into account a higher disk field strength and a more rapid disk
rotation closer to the planet the asymptotic velocity can be larger by 
a substantial factor.

\subsection{Comparison to simulations of MHD outflow formation}
In general,
two approaches have been undertaken to simulate the formation of
an MHD outflow.
One way is to treat the acceleration and collimation process {\em above}
the disk and consider the disk ``only'' as a boundary 
condition for the mass flow and the magnetic field structure fixed in time.
In the case of a monotonous disk magnetic field distribution 
a collimated outflow evolves out of a slow injection of matter in the
disk corona
(e.g. Ouyed \& Pudritz \cite{ouye97},
Fendt \& Cemeljic \cite{fend02}),
The other approach includes the evolution of the disk in the simulation
and the {\em disk-outflow connection}, i.e. the launching mechanism, can 
be investigated
(Hayashi et al.~\cite{haya96}; Miller \& Stone \cite{mill97};
Kuwabara et al.~\cite{kuwa00}).
However, mainly due to numerical problems with the accretion disk stability,
this approach does not yet allow to cover the long-term evolution of the
disk-outflow system.

All these simulations are performed considering normalized variables.
Having length scales normalized to an ``inner disk radius'',
velocities to the Keplerian velocity at this radius and
densities corresponding to a certain, given mass flow rate
(Ouyed \& Pudritz \cite{ouye97}),
the resulting simulations may be scaled from protostellar
outflow to the AGN, supposed that the derived parameters 
(as Mach number in the disk or outflow and disk temperature)
are consistent with the observations.
The question arises whether we can apply recent results of 
magnetohydrodynamic simulations of outflow formation 
(e.g. Ouyed \& Pudritz \cite{ouye97}; Fendt \& Elstner \cite{fend00};
Fendt \& Cemeljic \cite{fend02})
also for the model scenario of a protoplanet.

Essentially, there are three governing model assumptions involved in the
simulations, which are quite well known from the observations.
This is the fact that the accretion disk launching the outflow is geometrically
thin, implying that the sound speed is small compared to Keplerian speed.
Further, the outflow is ``cold'', i.e. gas pressure can be neglected
(low plasma beta).
The third assumption concerns the mass load of the outflow.
It is clear that we do not know these parameters for the case in the case
of planetary outflows due to the (yet existing) lack of observational evidence.
As a working hypothesis, we may assume so far that outflow formation works
similarly on all scales and that the MHD simulations of protostellar jet
formation can indeed be applied for the case of protoplanets.
The main results we can extrapolate are then
(i) asymptotic outflow velocity several times the Keplerian velocity
at the inner disk radius (Ouyed \& Pudritz \cite{ouye97}) and 
(ii) a rough equipartition between magnetic and kinetic energy in the
asymptotic flow.
For protoplanetary outflows launched from a disk radius of $10\,\rjup$
this would imply a asymptotic speed of at least $15\,{\rm km\,s^{-1}}$.

\section{Observational appearance of protoplanetary outflows}
Finally, we discuss possible observational features of protoplanetary 
outflows based on the model scenario developed earlier in this paper.
A Jovian mass protoplanetary core surrounded by a \cp accretion disk
orbits the central protostar along a Jupiter orbit.
Tidal interaction between the protoplanet and the \cs accretion disk
has opened-up a gap with in the \cs disk.
A large scale magnetic field below the equipartition field strength
is provided by the \cp accretion disk.
The outflow is launched magnetically from the \cp accretion disk.

\subsection{Outflow kinematics and geometry}
A critical measure for the outflow velocity are the escape velocities.
The escape velocity of a protoplanetary outflow from the gravitational 
potential of the protoplanet is
\begin{equation}
v_{\rm e,J}(r) = 59.6\,{\rm km\,s}^{-1} 
\left(\frac{r}{\rjup}\right)^{-1/2}
\left(\frac{M}{\Mjup}\right)^{1/2}.
\end{equation}
This value is similar to the lower limit for the asymptotic flow
velocity derived above for the Michel scaling, Eq.~\ref{eq_v_infty}.
In comparison, the escape speed for the protoplanetary outflow from the
gravitational potential of the {\em central star} at the orbit of Jupiter is
\begin{equation}
v_{\rm e,\star}(r) = 18.5\,{\rm km\,s}^{-1} 
\left(\frac{r}{D_{\rm J}}\right)^{-1/2}
\left(\frac{M}{\Msun}\right)^{1/2}.
\end{equation}
Therefore, a protoplanetary outflow propagating faster than the escape velocity
from the protoplanet will also be able to leave the gravitational potential of the
central star (see Fig.~1a).

For a standard thin {\em \cs} accretion disk of aspect ratio $h/r \simeq 0.1$,
the scale height at the orbit of Jupiter is $h \simeq 0.5\,AU \simeq 210\,\rjup$.
During the early phases of gap formation, the width of that gap will 
set an upper limit for the radius of the protoplanetary outflow.
%
If the outflow is launched during later stages of gap formation, 
MHD hoop-stresses may collimate the outflow to a radius smaller then the 
gap size.


During one orbital period a protoplanetary outflow would propagate 
``{\em length}'' of the outflow, is
\begin{equation}
l_{\rm out,J} = v_{\infty} P_{\rm orb,J} = 150\,AU
\left(\frac{v_{\infty}}{60\,{\rm km\,s}^{-1}}\right)
\end{equation}
(``{\em length}'' of the outflow).
This is a substantial distance if we consider the size of the outflow origin 
of only several protoplanetary radii.
If the life time of outflow is long enough (i.e. lasting more than several 
orbits),
we expect the overall structure of the outflow to be a {\em hollow tube} 
\footnote{Depending on the separation
of the planet from the central star, the outflow geometry may be characterized
better as a hollow {\em cone}.
Centrifugal imbalance with the central protostar will open up the tube geometry
into a cone as the outflow ascends from the ecliptic plane 
(see Fendt \& Zinnecker 1998, Masciadri \& Raga 2002 for
a similar discussion in the context of outflows from binary protostars)}
perpendicular to the \cs accretion disk
with a diameter of the order of the orbital radius.
The protoplanetary jet propagates within a layer along the tube/cone. 
The thickness of that layer is defined by the radius of the intrinsic outflow
and is probably several hundred planetary radii,
or about 1\% of the tube diameter.
Here we see a {\em major difference to protostellar jets} where the mass 
flow is distributed over the whole width of the global structure.

For an outflow velocity too low in order to leave the 
planetary gravitational potential, $v_{\rm esc,J} > v_{\rm out}$,
the material ejected from the disk will most probably be 
accumulated in a {\em blob} around the planet (see Fig.~1b).
As the blob remains bound to the planet, it will follow the orbit around
the central star.
The actual size of the blob will depend on the detailed balance between 
ejection of material out of the \cp disk and (spherical?) accretion of this
matter back to the protoplanetary system.
The maximum size of the blob is the Hill radius.
The possible existence of such an extended sphere of hot, shocked gas around 
the protoplanet is quite interesting from the observational point of view.

The kinetic luminosity
$P_{\rm kin} \simeq \dot{M}_{\rm out} v_{\rm out}^2$
of a protoplanetary outflow is about 
\begin{eqnarray}
P_{\rm kin} & \simeq & 10^{30}{\rm erg\,s}^{-1}
\left(\frac{\dot{M}_{\rm out}}{10^{-2}\dot{M}_{\rm cp}}\right)
\left(\frac{\dot{M}_{\rm cp}}{6\times 10^{-5}\,\Mjup\,{\rm yr}^{-1}}\right)
\nonumber \\
&\, & \quad\quad\quad \quad\quad
\cdot
\left(\frac{v_{\rm out}}{60\,{\rm km\,s}^{-1}}\right)^2
\end{eqnarray}
In comparison, for protostellar outflows with
$v_{\rm out} \simeq 500\,{\rm km\,s}^{-1}$ and 
$\dot{M}_{\rm out} \simeq 10^{-8} \Msun\,{\rm yr}^{-1}$
this value is four orders of magnitude higher.
%
We therefore conclude that a {\em protoplanetary outflow} could be detected
only if launched during a time period when no {\em protostellar outflow}
is present.
Otherwise the stellar outflow launched in the inner part of the \cs disk
and expanding outwards to about 50\,AU will just disrupt the narrow 
protoplanetary outflow.

An essential property of magnetic winds and outflows is the ability to
remove angular momentum very efficiently.
This is due to the large lever arm, which is defined by the
Alf\'en radius $r_A$ and typically a factor ten larger then
the foot point radius of the corresponding magnetic field line.
The angular momentum carried away per unit mass is
\begin{eqnarray}
L & = & 
r v_{\phi} - \frac{r B_p B_{\phi}}{4\pi \rho v_p} = \Omega_F r_A^2
\nonumber \\
 & = & 9.5\!\times\!10^{18} {\rm cm^2\,s^{-1}}
\left(\frac{M_p}{\Mjup}\right)^{1/2}
\!\!\left(\frac{r_F}{10\rjup}\right)^{1/2}
\!\!\left(\frac{r_A}{10 r_F}\right)^2
\end{eqnarray}
where $\Omega_F$ is the angular velocity at the foot point radius
$r_F$ of the field line.
The is a factor $(r_A/r_F)$ higher then in the pure hydrodynamic case with
$L \simeq \Omega_F r_F^2$.

The total angular momentum flux $\dot{J} \equiv \dot{M} L$
carried away by the protoplanetary outflow can be estimated as
\begin{eqnarray}
\dot{J}_{\rm out} & \simeq & 3\!\times\!10^{35} {\rm g\,cm^2\,s^{-2}}
\left(\frac{\dot{M}_{\rm out}}{10^{-2}\dot{M}_{\rm cp}}\right)
\left(\frac{\dot{M}_{\rm cp}}{6\times 10^{-5}\,\Mjup\,{\rm yr}}\right)
\nonumber \\
 & \, & 
\quad\quad
\quad\quad
\cdot 
\left(\frac{r_A}{10 r_F}\right)^2
\left(\frac{r_F}{10\rjup}\right)^{1/2}
\left(\frac{M_p}{\Mjup}\right)^{1/2}
\end{eqnarray}
With that, the total angular momentum loss over a orbital period at
a Jupiter distance is about 
$J_{\rm out} = 10^{44}{\rm g\,cm^2\,s^{-2}}$.

In comparison, the (rotational) angular momentum of Jupiter  
is $ J_{\rm J} = 6.3 \times 10^{45}{\rm g\,cm^2\,s^{-2}}$
(approximated by a homogeneous sphere).
Even if the mass carried away by the protoplanetary outflow is small
compared to the mass of the central planet, the angular momentum of
both components can be of similar order.
This leads to the conclusion that a planetary outflow launched for many
orbital periods of the protoplanet may well affect the angular momentum
evolution of the 
planetary core and, thus, the time scale for planet formation.

\subsection{Outflow density and observational features}
The time scale of the \cp accretion disk can be roughly estimated
to be of the order of 
\begin{equation}
\tau_{\rm disk} \simeq \Mjup /\dot{M}_{\rm cp} \simeq 2\times 10^4{\rm yr}
\simeq 1400\,P_{\rm orb,J}.
\end{equation}
Since the accretion rate decreases with time this value provides only
a lower limit.
It is clear, that the life time of the disk provides a natural upper limit for
the life time of the outflow\footnote{Statistical examinations as well as the
kinematic evolution of protostellar jets show that the life time of the outflow
must be considerably shorter than the disk time scale}.
The amount of mass deposited by the outflow into the ambient medium during its 
life time $\tau_{\rm out}$ is 
\begin{equation}
M_{\rm out} \simeq \dot{M}_{\rm out} \tau_{\rm out}
\simeq 
10^{-5}\Mjup\,
\left(\frac{\tau_{\rm out}}{10^{-3}\tau_{\rm disk}}\right)
\left(\frac{\dot{M}_{\rm out}}{10^{-2}\dot{M}_{\rm cp}}\right)
\end{equation}
On the other hand, 
from the outflow mass accumulated during one orbital period being confined in a
$100\,\rjup$-wide surface layer along the global outflow tube of 10\,AU 
diameter and 150\,AU length,
we derive mean a particle density within this layer of about
$2800\,{\rm cm}^{-3}$.
This value is surprisingly similar to protostellar outflows
(see e.g. Mundt et al.~\cite{mund90}).
As the velocities are not too different from protostellar jets, 
we therefore expect the existence of shock excited optical forbidden lines 
also in protoplanetary outflow.
The observed features would, however, look different.
In protostellar outflows we observe a wide shock structure extending over 
the whole 50 AU width of the outflow.
In comparison, in the case protoplanetary outflows one has to consider 
a bunch of ``shocklets'' distributed around the hollow tube of the flow.
As the kinetic luminosity of the proposed protoplanetary outflows is 
at least two orders of magnitude lower than for protostellar outflows,
we expect the same for luminosity in the shock emission.
However, as discussed above,
one can suspect that protostellar outflows are launched at times scales 
earlier than those considered the formation of protoplanets and their 
hypothetical outflows.

\subsection{Influence on the planet formation process}
Can protoplanetary outflows affect the process of planet formation?
Basically, the outflow restores accreted matter back into the ambient
medium while efficiently removing angular momentum from the accretion disk.

Considering the number values derived above we come to the conclusion 
that the mass evolution of the planet cannot be affected by the mass loss 
of the outflow as it is only about 1\% of the accretion rate. 

The impact of the enhanced angular momentum loss due to the magnetic outflow
can be twofold.
First, the {\em time-scale for the accretion process} can be decreased as the disk
angular momentum is removed more efficiently compared to purely viscous
transport in the disk.
The acceleration in the growth rate for the protoplanet is difficult to estimate.
A first guess would be that this time-scale is lowered by the same factor of
100 at which the angular momentum loss is increased.
Secondly, the angular momentum loss by the disk wind will influence the
{\em planet rotational period}. 
If angular momentum is redistributed into the interstellar medium and, thus,
not accreted by the planetary core, 
the final planet can be expected to rotate slower.

\section{Summary}
%
In this paper we have considered the possibility of launching 
magnetized outflows from \cp accretion disks.
We discuss a model scenario where the protoplanet is accompanied
by a circum-planetary accretion disk which is fed from the 
surrounding circum-stellar accretion disk.
This scenario is motivated by recent numerical simulations of planet formation.

For the outflow formation itself we suppose a mechanism similar to other
astrophysical outflows (protostellar jets, extragalactic jets) where the
flow is {\em magnetically} launched as a disk wind and then
accelerated and collimated into a narrow beam by Lorentz forces.

We have estimated the magnetic field strength of a fully convective
protoplanetary core and find an upper limit for the surface field
strength of about $10$\,G. 
For the surrounding \cp accretion disk we assume a accretion rate
of $\dot{M}_{\rm cp} = 6\times 10^{-5}\,\Mjup\,{\rm yr}$ for 
Jupiter-mass protoplanet.
The equipartition magnetic field strength in such a disk is
of a few $100\,$G and larger than the protoplanetary magnetic field.
The magnetic flux from the \cp accretion disk is estimated to
$5\times 10^{22}{\rm G\,cm}^2$.

We further investigated the \cp accretion disk temperature, its ionization
state and magnetic Reynolds number.
The \cp disk temperature may reach values up to $2000\,$K.
We find strong indication for a sufficient matter-field coupling 
underlining the magnetic character of the disk.
The latter is an essential condition for outflow launching.

From the estimated accretion rate and magnetic field strength we consider
the outflow magnetization for the asymptotic outflow velocity.
Applying the Michel scaling for magnetic outflows this velocity is about 
$63\,{\rm km\,s}^{-1}$ and of the order of the escape speed for the protoplanet.
However, a modified Michel scaling implied by the collimated structure
of the flow may result in even higher velocities.

For reasonable estimates for the outflow mass loss rate of 
$\dot{M}_{\rm out} = 10^{-2}\dot{M}_{\rm cp}$ and 
a velocity of
$v_{\rm out} = 60\,{\rm km\,s}^{-1}$,
the estimated kinetic power of a protoplanetary disk outflow is 
$ P_{\rm kin} \simeq 10^{30}{\rm erg\,s}^{-1}$.
These estimates also imply that during one orbit of the planet a mass
of about $10^{-5}$ Jupiter masses can be deposited in the interstellar
space by the protoplanetary outflow.
In the same time period this outflow would travel about $150$\,AU.

Two outflow scenarios are feasible depending on the outflow velocity.
If the outflow cannot escape the protoplanetary gravitational potential,
we expect the flow to build up a extended (bipolar) blob of hot
gas orbiting with the planet.
If the outflow exceeds the escape speed of the protoplanet, a collimated outflow
may be formed which (at a planetary orbit of about 5\,AU)
can also escape the stellar potential.

The overall outflow geometry built up by the orbiting source of a fast outflow 
is that of a hollow tube or hollow cone perpendicular to both \cp and \cs disk.
For the jet mass loss rate considered we find particle densities in the layer
along the tube of the outflow of $2800\,{\rm cm}^{-3}$,
a value similar to the density in protostellar jets.
Shock excited forbidden line emission can therefore be expected.

Energetically protoplanetary outflows cannot survive the interaction
with a protostellar jet which is launched from the inner regions
of the \cs accretion disk and can, thus, be only present if the stellar
outflow has ceased to exist.

The efficient angular momentum loss by the magnetized disk wind
may affect the accretion time scale and, thus, 
the time scale for planet formation.
The angular momentum loss by the disk wind or outflow per one orbital 
period is about the rotational angular momentum of today's Jupiter.

In summary, our model estimates strongly rely on the accretion rate of the 
\cp accretion disk.
We apply $\dot{M}_{\rm cp} = 6\times 10^{-5}\,\Mjup\,{\rm yr}$, a value
which is supported by many independent numerical simulations of planet
formation in accretion disks.
Further model constraints can be expected from future, high resolution numerical
simulations of the \cp accretion disk clarifying the detailed \cp disk structure
close to the protoplanet.

\begin{acknowledgements}
The author thanks Pavel Kroupa for directing his interest to the question
of protoplanetary outflows. 
The present work has benefited from discussions with Detlef Elstner,
Willy Kley, Matthias Steffen and, in particular, Gennaro D'Angelo.
Hans Zinnecker and Joachim Wambsganss are acknowledged for their 
continuous support.
This work has been financed by the research grant 
HSPN 24-04/302;2000 of the State of Brandenburg.
\end{acknowledgements}

\begin{appendix}

\section{The standard disk parameters}
Here we apply the Shakura \& Sunyaev (\cite{shak73}) accretion disk parametrisation
of a thin $\alpha$-viscosity disk to the case of \cp disks. 
We show the temperature and density profile and for comparison the same also
for the \cs disk. 
In our case radiation pressure can be neglected. Free-free absorption dominates
electron scattering. The sound speed is given by 
$c_{\rm s} = (k_{\rm B} T/m_{\rm prot})$.
%
%
The particle density is
\begin{eqnarray}
n(r) 
& = & 1.6\times 10^{14}{\rm cm}^{-3} \alpha^{-7/10}
\left(\frac{\dot{M}}{6\times 10^{-5}\Mjup {\rm yr}^{-1}}\right)^{11/20}
\nonumber\\
& \, &
\quad \quad
\quad \quad
\quad \quad
\quad \quad
\cdot \left(\frac{M}{\Mjup}\right)^{5/8}
\left(\frac{r}{15\,\rjup}\right)^{-15/8} \\
& = & 2.8\times 10^{14}{\rm cm}^{-3} \alpha^{-7/10}
\left(\frac{\dot{M}}{1.2\times 10^{-7}\Msun {\rm yr}^{-1}}\right)^{11/20}
\nonumber\\
&\, & 
\quad \quad
\quad \quad
\quad \quad
\quad \quad
\cdot \left(\frac{M}{\Msun}\right)^{5/8}
\left(\frac{r}{15\,\rsun}\right)^{-15/8}
\end{eqnarray}
while the disk temperature is 
\begin{eqnarray}
T(r) 
& = & 2800 {\rm K}\,\alpha^{-1/5}
\left(\frac{\dot{M}}{6\times 10^{-5}\Mjup {\rm yr}^{-1}}\right)^{3/10}
\nonumber\\
& \, &
\quad \quad
\quad \quad
\quad \quad
\quad \quad
\cdot \left(\frac{M}{\Mjup}\right)^{1/4}
\left(\frac{r}{15\,\rjup}\right)^{-3/4}\\
& = & 3700 {\rm K}\,\alpha^{-1/5}
\left(\frac{\dot{M}}{1.2\times 10^{-7}\Msun {\rm yr}^{-1}}\right)^{3/10}
\nonumber\\
&\, & 
\quad \quad
\quad \quad
\quad \quad
\quad \quad
\cdot \left(\frac{M}{\Msun}\right)^{1/4}
\left(\frac{r}{15\,\rsun}\right)^{-3/4}
\end{eqnarray}
The temperature derived here is the mid-plane disk temperature and 
is in agreement with the simple estimate of the surface temperature
in Eq.~\ref{eq_t_disk}, but also the more sophisticated disk model 
probing the vertical \cs disk structure (D'Alessio et al.~\cite{dale98}).
We note, however, 
that in spite of the general success of the Shakura \& Sunyaev 
$\alpha$-parametrization of turbulence, in particular for hot disks
around compact objects,
it is known that the this model does not fit very well all of the observed 
disks of T\,Tauri stars (i.e. cool disks), in which a $T(r) \sim r^{-1/2}$ 
temperature profile indicated 
(Beckwith et al.~\cite{beck90}).

Finally, we supply the expression for the disk-half thickness from
Shakura \& Sunyaev for the case of a circum-planetary disk,
\begin{eqnarray}
\label{eq_hr}
h(r) 
& = & 6.5\times10^6{\rm cm}\,\alpha^{-1/10}
\left(\frac{\dot{M}}{6\times 10^{-5}\Mjup {\rm yr}^{-1}}\right)^{3/20}
\nonumber\\
& \, &
\quad \quad
\quad \quad
\quad \quad
\quad \quad
\cdot \left(\frac{M}{\Mjup}\right)^{9/10}
\left(\frac{r}{15\,\rjup}\right)^{9/8}.
\end{eqnarray}
Compared to the \cs accretion disk, the \cp is expected to somewhat thinner.

\end{appendix}


\end{document}